\documentclass[11pt]{article}
\pdfoutput=1

\usepackage[top=3cm, bottom=2cm, left=3.5cm, right=3.5cm, includefoot]{geometry}

\usepackage{enumitem}
\usepackage{color}

\usepackage{graphicx}
\usepackage{epstopdf}
\usepackage[ruled,linesnumbered,algo2e,vlined]{algorithm2e}
\usepackage{multirow}
\usepackage[flushleft]{threeparttable}
\usepackage{authblk}

\SetCommentSty{mycommfont}

\newtheorem{theorem}{Theorem}
\newtheorem{lemma}{Lemma}
\newtheorem{definition}{Definition}
\newtheorem{example}{Example}
\newtheorem{problem}{Problem}
\newtheorem{remark}{Remark}

\usepackage{amssymb}
\usepackage{amsmath}

\SetKw{Continue}{continue}

\newcommand{\etal}{\textit{et~al.}}

\newcommand{\exfmatchfull}{function match with variables-to-constants mapping}
\newcommand{\exfmatchingfull}{function matching problem with variables-to-constants mapping}

\newcommand{\exfmatch}{FVC-match}
\newcommand{\exfmatches}{FVC-matches}
\newcommand{\exfmatching}{FVC-matching}

\newcommand{\expmatchfull}{parameterized match with variables-to-constants mapping}
\newcommand{\expmatchingfull}{parameterized pattern matching problem with variables-to-constants mapping}

\newcommand{\expmatch}{PVC-match}
\newcommand{\expmatches}{PVC-matches}
\newcommand{\expmatching}{PVC-matching}

\newcommand{\concat}{}
\newcommand{\emptystr}{\epsilon}
\newcommand{\conv}{\otimes}

\newcommand{\dontcare}{{\tt \ast}}

\newcommand{\piapply}[1]{\hat{#1}}

\newcommand{\apeq}{\lower0.4ex\hbox{{$\;\stackrel{\bowtie}{=}\;$}}}

\newcommand{\apneq}{\lower0.4ex\hbox{{$\;\not\stackrel{\bowtie}{=}\;$}}}

\newcommand{\isvar}[1]{v(#1)}
\newcommand{\tta}{{\tt a}}
\newcommand{\ttb}{{\tt b}}
\newcommand{\ttc}{{\tt c}}
\newcommand{\ttA}{{\tt A}}
\newcommand{\ttB}{{\tt B}}
\newcommand{\ttC}{{\tt C}}

\newcommand{\Paster}{P_{\!\!\dontcare}}
\newcommand{\Pvar}{P_{\!\Pi}}
\newcommand{\Tvar}[1]{T'_{#1}}
\newcommand{\Tvarsingle}{T'}
\newcommand{\TT}{{\textit{\textbf{T}}}}

\newcommand{\Pvarx}{P_x}

\newcommand{\mtt}[1]{\mathtt{#1}}

\title{New Variants of Pattern Matching with Constants and Variables}

\author[]{Yuki Igarashi}
\author[]{Diptarama}
\author[]{Ryo Yoshinaka}
\author[]{Ayumi Shinohara}

\affil[]{Graduate School of Information Sciences, Tohoku University, Sendai, Japan \newline
	\texttt{\{yuki\_igarashi, diptarama\}@shino.ecei.tohoku.ac.jp \newline 
		\{ry, ayumi\}@ecei.tohoku.ac.jp}}

\begin{document}

\maketitle

\begin{abstract}
	Given a text and a pattern over two types of symbols called constants and variables,
the \emph{parameterized pattern matching problem} is to find all occurrences of substrings of the text that the pattern matches by substituting a variable in the text for each variable in the pattern, where the substitution should be injective.
The \emph{function matching problem} is a variant of it that lifts the injection constraint.
In this paper, we discuss variants of those problems, where one can substitute a constant or a variable for each variable of the pattern.
We give two kinds of algorithms for both problems, a convolution-based method and an extended KMP-based method,
and analyze their complexity.
\end{abstract}

\section{Introduction}

The \emph{parameterized pattern matching} problem was proposed by Baker~\cite{ref:baker1996parameterized} about a quarter of a century ago.
Problem instances are two strings called a pattern and a text, which are sequences of two types of symbols called constants and variables.
The problem is to find all occurrences of substrings of a given text that a given pattern matches
by substituting a variable in the text for each variable in the pattern, where
the important constraint is that the substitution should be an injective map.
She presented an algorithm for this problem that runs in $O(n\log{n})$ time using \emph{parameterized suffix trees}, where $n$ is the length of text.

By removing the injective constraint from the parameterized pattern matching problem, Amir \etal{}~\cite{ref:amir2006function} proposed the \emph{function matching} problem, where the same variable may be substituted for different variables.
Yet another but an inessential difference between parameterized pattern matching and function matching is in the alphabets.  The function matching problem is defined to be constant-free in the sense that patterns and texts are strings over variables.  However, this simplification is inessential, since it is known that the problem with variables and constants is linear-time reducible to the constant-free case~\cite{ref:amir1994alphabet}.  This reduction technique works for the parameterized pattern matching as well. 
Their deterministic algorithm solves this problem in $O(|\Pi|n\log{m})$ time, where $n$ and $m$ are the lengths of the text and pattern, respectively, and $|\Pi|$ is the number of different symbols in the pattern.
After that, Amir and Nor~\cite{ref:amir2007generalized} introduced the \emph{generalized function matching} problem, where one can substitute a string of arbitrary length for a variable.
In addition, both a pattern and a text may contain ``don't care'' symbols, which are supposed to match arbitrary strings.

The parameterized pattern matching problem and its extensions have been great interests not only to the pattern matching community~\cite{ref:mendivelso2015parameterized} but also to the database community.
Du Mouza \etal{}~\cite{ref:du2007ppatternqueries} proposed a variant of the function matching problem, where texts should consist solely of constants and a substitution maps variables to constants, which is not necessarily injective.
Let us call their problem \emph{function matching with variables-to-constants, FVC-matching} in short.\footnote{They called the problem \emph{parameterized pattern queries}. However, to avoid misunderstanding the problem to have the injective constraint, we refrain from using the original name in this paper.}
The function matching problem is linear-time reducible to this problem by simply assuming the variables in a text as constants.
Therefore, this problem can be seen as a generalization of the function matching problem.
Unfortunately, as we will discuss in this paper, their algorithm is in error.

In this paper, we introduce a new variant of the problem by du Mouza \etal{} with the injective constraint, which we call \emph{parameterized pattern matching with variables-to-constants mapping (PVC-matching)}.
For each of the FVC-matching and PVC-matching problems, we propose two kinds of algorithms\footnote{Source codes for those algorithms are available at \\\texttt{https://github.com/igarashi/matchingwithvcmap}.}: a \emph{convolution-based method} and an \emph{extended KMP-based method}.
The convolution-based methods and extended KMP-based methods are inspired by the algorithm of Amir \etal{}~\cite{ref:amir2006function} for the function matching problem and the one by du Mouza \etal{}~\cite{ref:du2007ppatternqueries} for the FVC-matching problem, respectively.
As a result, we fix the flaw of the algorithm by du Mouza \etal{}
The convolution-based methods for both problems run in $O(|\Sigma_{P}|n\log{m})$ time, where $\Sigma_P$ is the set of constant symbols that occur in the pattern $P$.
Our KMP-based methods solve the PVC-matching and FVC-matching problems with $O(|\Pi|(|\Sigma_{P}|+|\Pi|)m^2)$ and $O(|\Pi|n)$) preprocessing time and $O(|\Pi| \lceil \frac{m}{w} \rceil n)$ and $O(|\Pi_{P}|^2 \lceil \frac{m}{w} \rceil n)$ query time, respectively, where
 $\Pi$ is the set of variables and $w$ is the word size of a machine (Table~\ref{fig:tcomplexity}).
\begin{table}[tb]
	\centering
	\caption{The time complexity of our proposed algorithms}
	\label{fig:tcomplexity}
\scalebox{1}[1]{
\begin{tabular}{|c|c|c|c|}
	\hline
	\multirow{2}{*}{Problem} & \multirow{2}{*}{\begin{tabular}[c]{@{}c@{}}Convolution-based\\ Method\end{tabular}} & \multicolumn{2}{c|}{Extended KMP-based Method} \\ \cline{3-4} 
	&  & Preprocessing & Query \\ \hline
	\expmatching{} & \multirow{2}{*}{$O(|\Sigma_{P}|n\log{m})$} & $O(|\Pi_{P}||\Sigma_{P}|m^2)$ & $O(|\Pi_{P}|\lceil \frac{m}{w} \rceil n)$ \\ \cline{1-1} \cline{3-4} 
	\exfmatching{} &  & $O(|\Pi_{P}|(|\Sigma_{P}|+|\Pi_{P}|)m^2)$ & $O(|\Pi_{P}|^2 \lceil \frac{m}{w} \rceil n)$ \\ \hline
\end{tabular}
}
\end{table}

\section{Preliminaries}

For any set $Z$, the cardinality of $Z$ is denoted by $|Z|$.
Let $\Sigma$ be an alphabet.
We denote by $\Sigma^*$ the set of strings over $\Sigma$.
The empty string is denoted by $\emptystr$.
The concatenation of two strings $X,Y \in \Sigma^*$ is denoted by $X\concat Y$.
For a string $X$, the length of $X = X[1]\concat X[2]\concat \cdots \concat X[n]$ is denoted by $|X| = n$.
The substring of $X$ beginning at $i$ and ending at $j$ is denoted by $X[i:j] = X[i]\concat X[i+1]\concat \cdots X[j-1]\concat X[j]$.
Any substrings of the form $X[1:j]$ and $X[i:n]$ are called a \emph{prefix} and a \emph{suffix} of $X$.
For any number $ k $, we define $X[k:k-1]=\emptystr$.
The set of symbols from a subset $\Delta$ of $\Sigma$ occurring in $X$ is denoted by $\Delta_X = \{\, X[i] \in \Delta \mid 1 \le i \le n \,\}$.

This paper is concerned with matching problems, where strings consist of two kinds of symbols, called \emph{constants} and \emph{variables}.
Throughout this paper, the sets of constants and variables are denoted by $\Sigma$ and $\Pi$, respectively.
Variables are supposed to be replaced by another symbol, while constants are not.
\begin{definition}\label{def:funcapplication}
	For a function $\pi: \Pi \to (\Sigma \cup \Pi)$, we extend it to $\hat{\pi}: (\Pi \cup \Sigma)^* \to (\Pi \cup \Sigma)^*$ by
	\begin{eqnarray*}
		\piapply{\pi}(X) = \hat{\pi}(X[1])\concat \hat{\pi}(X[2]) \concat \cdots \concat \hat{\pi}(X[n]) , 
		\text{where\space} \piapply{\pi}(X[i]) = \begin{cases}
			\pi(X[i]) & (X[i] \in \Pi) \\
			X[i] & {\rm (otherwise)}
		\end{cases}
	\end{eqnarray*}	
\end{definition}

Parameterized match~\cite{ref:baker1996parameterized} and function match~\cite{ref:amir2006function}\footnote{Amir \etal{}~\cite{ref:amir2006function} defined the problem so that strings are over a single type of symbols, which can be seen as variables. This restriction is inessential~\cite{ref:amir1994alphabet}.} are defined as follows.
\begin{definition}
	Let $P$ and $Q$ be strings over $\Sigma \cup \Pi$ of the same length.
	String $P$ is said to \emph{parameterized match} (resp.\ \emph{function match})
	  string $Q$ if there exists an injection (resp.\ function) $\pi: \Pi \to \Pi$, such that $\piapply{\pi}(P) = Q$.
\end{definition}
The parameterized pattern matching problem (resp. function matching problem) is to find all occurrences of substrings of a given text that a given pattern parameterized match (resp.\ function match).

The problems we discuss in this paper allow variables to be mapped to constants and variables.
\begin{definition}\label{def:proposedproblem}
	Let $P$ and $Q$ be strings over $\Sigma \cup \Pi$ of the same length.
	String $P$ is said to \emph{\expmatchfull{}} (resp. \emph{\exfmatchfull{}}), shortly \emph{\expmatch{}} (resp. \emph{\exfmatch{}}), string $Q$ if there exists an injection (resp. function) $\pi: \Pi \to (\Sigma \cup \Pi)$, such that $\piapply{\pi}(P) = Q$.
\end{definition}
\begin{problem}
	Let $P$ and $T$ be strings over $\Sigma \cup \Pi$ of length $m$ and $n$, respectively.
	The \emph{\expmatchingfull{}} (resp. \emph{\exfmatchingfull{}}), shortly \emph{\expmatching{}} (resp. \emph{\exfmatching{}}) asks for all the indices $i$ where pattern $P$ \emph{\expmatches{}} (resp. \emph{\exfmatches{}}) substring $T[i: i+m-1]$ of text $T$.
\end{problem}
Table~\ref{fig:problemdef} summarizes those four problems.
\begin{table}[tb]
	\centering
	\caption{Definition of problems}
	\label{fig:problemdef}
\begin{tabular}{|rc|c|c|}
	\hline
	\multirow{2}{*}{Problems} & & \multicolumn{2}{|c|}{Admissible mappings}
\\ \cline{3-4}
	& &  Type & Injection constraint   \\ \hline
	\expmatching{} &  & \multirow{2}{*}{$\Pi \to (\Pi \cup \Sigma)$} & Yes \\ \cline{1-2} \cline{4-4} 
	\exfmatching{} & \cite{ref:du2007ppatternqueries} &  & No  \\ \hline
parameterized matching & \cite{ref:baker1996parameterized}  & \multirow{2}{*}{$\Pi \to \Pi$} & Yes \\ \cline{1-2} \cline{4-4} 
function matching  & \cite{ref:amir2006function} & & No \\ \hline
\end{tabular}
\end{table}

We can assume without loss of generality that the text $T$ solely consists of constants.
This restriction is inessential since one can regard variables occurring in $T$ as constants.
Under this assumption, the FVC-matching problem is exactly \emph{parameterized pattern queries}~\cite{ref:du2007ppatternqueries}.

\begin{example} \label{example:matching}
	Let $\Sigma = \{\texttt{a}, \texttt{b}\}$ and $\Pi = \{\texttt{A}, \texttt{B}\}$.
	Consider pattern $P = \texttt{ABAb}$ and text $T = \texttt{ababbbb}$.
	Then, the answer of \expmatching{} problem is $\{1, 2\}$, since $P$ \expmatches{} $T[1:4] = \texttt{abab}$, $T[2:5] = \texttt{babb}$.
	On the other hand, the answer of \exfmatching{} problem is $\{1, 2, 4\}$ since $P$ \exfmatches{} $T[1:4] = \texttt{abab}$, $T[2:5] = \texttt{babb}$, $T[4:7] = \texttt{bbbb}$.
	Note that we have $\hat{\pi}(P)=T[4:7]$ for $\pi$ with $\pi(\mtt{A})=\pi(\mtt{B})=\mtt{b}$, which is not injective.
\end{example}
Throughout this paper, we arbitrarily fix a pattern $P \in (\Sigma \cup \Pi)^*$ of length $m$ and a text $T \in \Sigma ^*$ of length $n$.
\section{Convolution-based Methods}

In this section, we show that the \exfmatching{} problem can be solved in $O(|\Sigma_{P}|n\log{m})$ time by reducing the problem to the function matching problem and the wildcard matching problem, for which several efficient algorithms are known.
The \expmatching{} problem can also be solved using the same reduction technique with a slight modification.

For strings $P$ of length $m$ over $\Sigma \cup \Pi$ and $T$ of length $n$ over $\Sigma$, we define $\Pi' = \Pi_{P} \cup \Sigma_{T}$.
Let $\Paster \in (\Sigma \cup \{\dontcare\})^*$ be a string obtained from $P$ by replacing all variable symbols in $\Pi$ with \emph{don't care symbol} $\dontcare$.
Let $\Pvar \in \Pi'^*$ be a string obtained from $P$ by removing all constant symbols in $\Sigma$.
Moreover, for $1 \le i < n-m$, let $\Tvar{i}$ be a string defined by $\Tvar{i} = \isvar{1} \isvar{2} \concat \cdots \concat \isvar{m}$, where $\isvar{j} = T[i+j-1]$ if $P[j] \in \Pi$ and $\isvar{j} = \emptystr$ otherwise.
Note that both the lengths of $\Tvar{i}$ and $\Pvar$ are equal to the total number of variable occurrences in $P$.
\begin{example}
	For $T = {\tt aabcbc}$ and $P={\ttA \tta \ttB \ttB \ttb}$ over $\Pi = \{\ttA,\ttB\}$ and $\Sigma = \{\tta,\ttb,\ttc\}$,
	we have 
	$\Paster = \dontcare \tta \dontcare \dontcare \ttb$, 
	$\Pvar=\ttA\ttB\ttB$, 
	$\Tvar1 = \tta \ttb \ttc$, and 
	$\Tvar2= \tta \ttc \ttb$.
\end{example}

For both \exfmatching{} and \expmatching{} problems, the following lemma is useful to develop algorithms to solve them.
\begin{lemma} \label{obs:fftreduction}
	$P$~\text{\exfmatches{} (resp.\ \expmatches{})} $T[i:i+m-1]$ if and only if
	\begin{enumerate}[topsep=0pt]
		\item $\Paster$ wildcard matches $T[i:i+m-1]$, and
		\item $\Pvar$ function matches (resp.\ parameterized matches) $\Tvar{i}$.
	\end{enumerate}
\end{lemma}
Lemma~\ref{obs:fftreduction} suggests that the FVC-problem would be reducible to the combination of wildcard matching problem and function matching problem.

The wildcard matching problem~(a.k.a.\ Pattern matching with don't care symbol)~\cite{ref:fischer1974string} is one of the fundamental problems in pattern matching.
There are many algorithms for solving the wildcard matching problem.
Fischer \etal~\cite{ref:fischer1974string} gave an algorithm for (a generalization of) this problem, which runs in $O({|\Sigma|}n\log{m})$ time.
Cole and Hariharan~\cite{ref:cole2002verifying} improved it to $O(n\log{m})$ time by using convolution.
On the other hand, Pinter~\cite{ref:pinter1985efficient} gave an $O(n + m + \alpha)$-time algorithm,
where $\alpha$ is the total number of occurrences of the maximal consecutive constant substrings of the pattern in the text.
This algorithm uses Aho-Corasick algorithm instead of convolution.
Iliopoulos and Rahman~\cite{ref:iliopoulos2007pattern} proposed an algorithm which utilizes suffix arrays for text.
The algorithm preprocesses a text in $O(n)$ time and runs in $O(m+\alpha)$ time.

However, Lemma~\ref{obs:fftreduction} does \emph{not} imply the existence of a \emph{single} string $\Tvarsingle$ such that $P$ FVC-matches $T[i:i+m-1]$ 
 if and only if $\Paster$ wildcard matches $T[i:i+m-1]$ and 
 $\Pvar$ function matches $\Tvarsingle[i:i+m-1]$.
A naive application of Lemma~\ref{obs:fftreduction} to compute $\Tvar{i}$ explicitly for each $i$
requires $O(mn)$ time in total.

We will present an algorithm to check whether $\Pvar$ function matches (parameterized matches) $\Tvar{i}$ for all $1 \le i < n-m$ in $O(\log{|\Sigma|}\,n\log{m})$ time in total.
Without loss of generality, we assume that $\Sigma$ and $\Pi$ are disjoint finite sets of positive integers in this section, and for integers $a$ and $b$, the notation $a \cdot b$ represents the multiplication of $a$ and $b$ but not the concatenation.
\begin{definition}\label{def:convolution}
	For integer arrays $A$ of length $n$ and $B$ of length $m$, we define an integer array $R$ by
\(
		R[j] = \sum_{i=1}^{m}{A[i+j-1]\cdot B[i]}
\)	
	for $1 \le j \le n-m+1$.
	We denote $R$ as $A\conv B$.
\end{definition}
In a computational model with word size $O(\log{m})$, the discrete convolution can be computed in time $O(n\log{n})$ by using the Fast Fourier Transform (FFT)~\cite{gormen1990introduction}.
The array $R$ defined in Definition \ref{def:convolution} can also be computed in the same time complexity by just reversing array $B$.

Amir \etal{}~\cite{ref:amir2006function} proved the next lemma for function matching.
\begin{lemma}[\cite{ref:amir2006function}]\label{lem:amirlemma}
	For any natural numbers $a_{1},\cdots,a_{k}$,
the equation \\
$k\!\cdot\!\sum_{i = 1}^{k}{(a_i)^2} = ( \sum_{i=1}^{k}{a_i} )^2$ holds
if and only if $a_i = a_j \mbox{ for any } 1 \le i, j \le k$.
\end{lemma}

Let $\TT$ be the string of length $n$ such that $\TT[i] = (T[i])^2$ for every $1 \leq i \leq n$.
For a variable $x \in \Pi_{P}$, 
let $c_{x}$ denote the number of occurrences of $x$ in $P$, and 
let $\Pvarx$ be the string of length $m$ such that $\Pvarx[j] = 1$ if $P[j] = x$ and $\Pvarx[j] = 0$ otherwise, for every $1 \le j \le m$.
By Lemma~\ref{lem:amirlemma}, we can prove the following lemma.
\begin{lemma} \label{lemma:uniqueSymbol}
	All the symbols (values) in $\Tvar{i}$ at every position $j$ satisfying $\Pvar[j] = x$ are the same,
    if and only if the equation
	$c_{x} \!\cdot\! ((\TT \conv \Pvarx)[i]) = ((T \conv \Pvarx)[i])^2$ holds.  
\end{lemma}
Thus, $\Pvar$ function matches $\Tvar{i}$ if and only if the equation in Lemma~\ref{lemma:uniqueSymbol} holds for all $x \in \Pi_{P}$.
Both the convolutions $\TT \conv \Pvarx$ and $T \conv \Pvarx$  can be calculated in $O(n\log{m})$ time by simply dividing $T$ into $2\times\frac{n}{2m}$ overlapping substrings of length $2m$.
For parameterized pattern matching problem, we have only to check additionally whether 
the value ${(T \conv \Pvarx)[i]}/c_{x}$ is unique among all $x \in \Pi_{P}$.
A pseudo code for solving the \expmatching{} problem using convolution is shown as Algorithm~\ref{alg:fftbased} in Appendix~\ref{sec:appendixalgorithms}.

\begin{theorem}
	The \exfmatching{} problem and \expmatching{} problem can be solved in $O(|\Sigma_{P}|\,n\log{m})$ time.
\end{theorem}

\section{KMP-based Methods}
Du Mouza \etal{} proposed a KMP-based algorithm for the \exfmatching{} problem, which, however, is in error.
In this section, we propose a correction of their algorithm, which runs in $O(|\Pi|^2  \lceil \frac{m}{w} \rceil n)$ query time with $O(|\Pi|(|\Sigma_{P}|+|\Pi|)m^2)$ preprocessing time,
 where $w$ denotes the word size of a machine.
This algorithm will be modified so that it solves the \expmatching{} problem in $O(|\Pi|  \lceil \frac{m}{w} \rceil n)$ query time with $O(|\Pi||\Sigma_{P}|m^2)$ preprocessing time.

The KMP algorithm~\cite{ref:knuth1977fast} solves the standard pattern matching problem in $O(n)$ time with $O(m)$ preprocessing time.
We say that a string $Y$ is a \emph{border} of $X$ if $Y$ is simultaneously a prefix and a suffix of $X$.  A border $Y$ is \emph{nontrivial} if $Y$ is not $X$ itself.
For the preprocessing of the KMP algorithm, we calculate the longest nontrivial border $b_{k}$ for each prefix $P[1:k]$ of pattern $P$, and store them as \emph{border array} $B[k] = |b_{k}|$ for each $0 \le k \le m$.
Note that $b_0 = b_1 = \epsilon$.
In the matching phase, the KMP algorithm compares symbols $T[i]$ and $P[k]$ from $i = k = 1$.
We increment $i$ and $k$ if $T[i] = P[k]$.
Otherwise we reset the index for $P$ to be $k' = B[k-1]+1$ and resume comparison from $T[i]$ and $P[k']$.

\subsection{Extended KMP Algorithm}
This subsection discusses an algorithm for the FVC-matching problem.
In the matching phase, our extended KMP algorithm compares the pattern and a substring of the text in the same manner as the classical KMP algorithm
 except that we must maintain a function by which prefixes of the pattern match some substrings of the text.
That is, our extended KMP algorithm compares symbols $T[i]$ and $P[k]$ from $i = k = 1$ with the empty function $\pi$.
If $P[k]$ is not in the domain $\mathrm{dom}(\hat{\pi})$ of $\hat{\pi}$, we expand $\pi$ by letting $\pi(P[k]) = T[i]$ and increment $i$ and $k$.
If $\hat{\pi}(P[k])$ is defined to be $T[i]$, we increment $i$ and $k$.
Otherwise, we say that \emph{a mismatch occurs at position $k$ with a function $\pi$}.
Note that the mismatch position refers to that of $P$ rather than $T$.
When we find a mismatch, we must calculate the appropriate position $j$ of $P$ and function $\pi'$ with which we resume comparison.
If instances are variable-free, the position is solely determined by the longest border size of $P[1:k]$ and we have no function.
In the case of FVC-matching, the resuming position depends on the function $\pi$ in addition to $k$.
\begin{example}\label{ex:exkmp}
Let us consider the pattern $P=\mathtt{AABaaCbC}$ where $\Pi = \{\ttA,\ttB,\ttC\}$ and $\Sigma = \{\tta,\ttb\}$ in Fig.~\ref{fig:kmpexkmp}.
If the concerned substring of the text is $T'=\mathtt{bbbaaabb}$, a mismatch occurs at $k=8$ with a function $\pi$ such that $\pi(\mtt{A})=\pi(\mtt{B})=\mtt{b}$ and $\pi(\mtt{C})=\mtt{a}$.
In this case, we can resume comparison with $P[7]$ and $T'[8]$, since we have $\hat{\pi}'(P[1:6])=T'[2:7]$ for $\pi'$ such that $\pi'(\mtt{A})=\pi'(\mtt{C})=\mtt{b}$ and $\pi'(\mtt{B})=\mtt{a}$.
On the other hand, for $T''= \mtt{bbaaaabb}$, the first mismatch occurs again at $k=8$ with a function $\rho$ such that $\rho(\mtt{A})=\mtt{b}$ and $\rho(\mtt{B})=\rho(\mtt{C})=\mtt{a}$.
In this case, one cannot resume comparison with $P[7]$ and $T''[8]$, since there is no $\rho'$ such that $\hat{\rho}'(P[1:6])=T''[2:7]$, since $P[1] = P[2]$ but $T''[2] \neq T''[3]$.
We should resume comparison between $P[4]$ and $T''[8]$ with $\rho'$ such that $\rho'(\mtt{A})=\mtt{a}$ and $\rho'(\mtt{B})=\mtt{b}$, for which we have $\hat{\rho}'(P[1:3])=T''[5:7]$.
Note that $\rho'(\mtt{C})$ is undefined.
\begin{figure}[t]
	\centering
		\centering
		\includegraphics[width=0.7\textwidth]{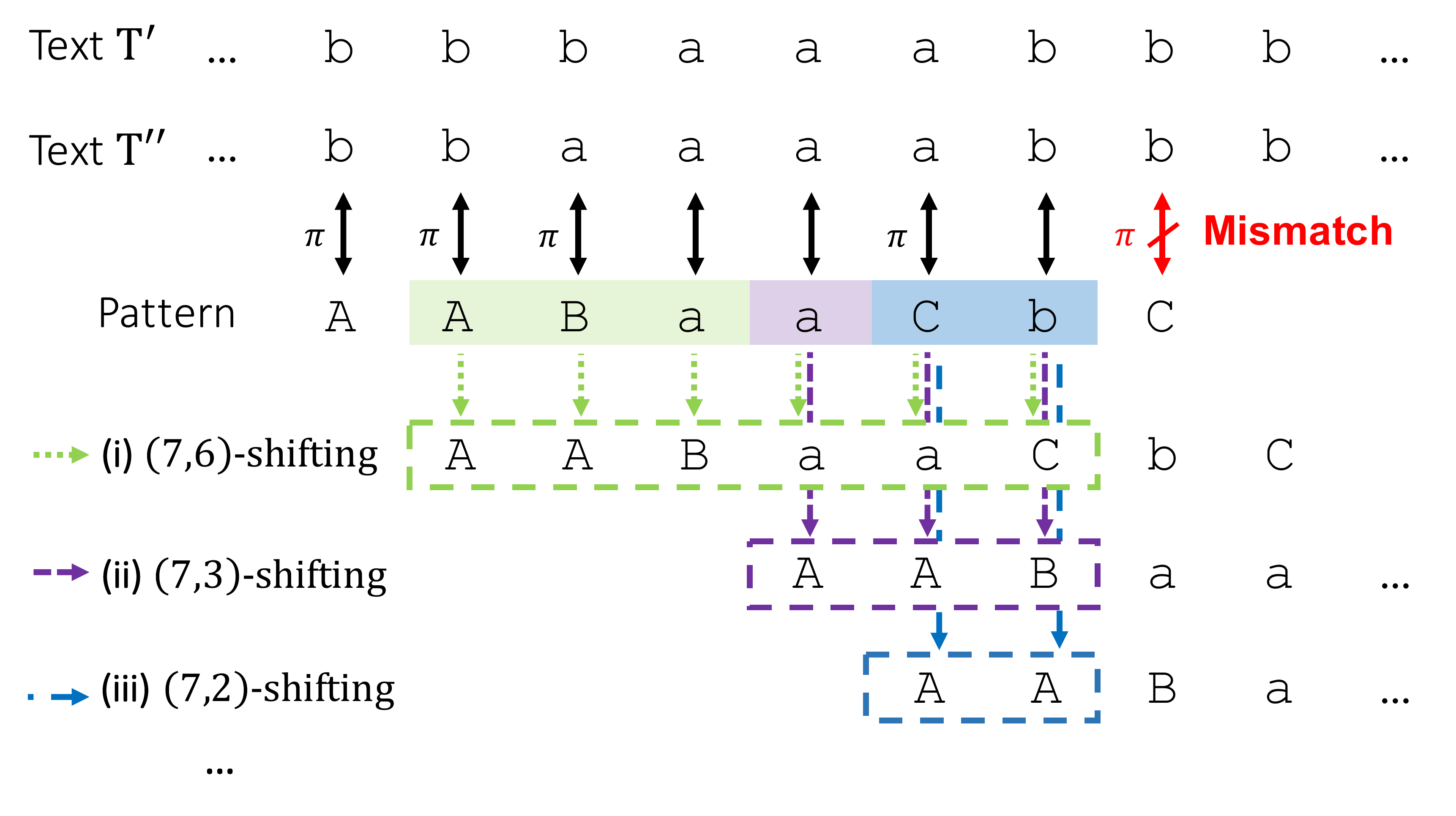}
	\caption{Examples of possible shifts in the Extended KMP algorithm}
	\label{fig:kmpexkmp}
\end{figure}
\end{example}

The goal of the preprocessing phase is to prepare a data structure by which one can efficiently compute the \emph{failure function} in the matching phase:
\begin{itemize}
	\item[] \textbf{Input:} the position $k+1$ (where a mismatch occurs) and a function $\pi$ whose domain is $\Pi_{P[1:k]}$,
	\item[] \textbf{Output:} the largest position $j+1 < k+1$ (at which we will resume comparison) and the function $\pi'$ with domain $\Pi_{P[1:j]}$ such that $\hat{\pi}'(P[1:j]) = \hat{\pi}(P[k-j+1:k])$.
\end{itemize}
	We call such $\pi$ a \emph{preceding function}, $\pi'$ a \emph{succeeding function} and the pair $(\pi,\pi')$ a \emph{$(k,j)$-shifting function pair}.
	The substrings $P[1:j]$ and $P[k-j+1:k]$ may not be a border of $P[1:k]$ but under preceding and succeeding functions they play the same role as a border plays in the classical KMP algorithm.
	The succeeding function $\pi'$ is uniquely determined by a preceding function $\pi$ and positions $k,j$.
	The condition that functions $\pi$ and $\pi'$ form a $(k,j)$-shifting function pair can be expressed using the \emph{$(k,j)$-shifting graph} (on $P$), defined as follows.
	
\begin{figure}[h!]
	\centering
	\begin{minipage}[t]{0.36\hsize}
	\centering
	\includegraphics[width=\textwidth]{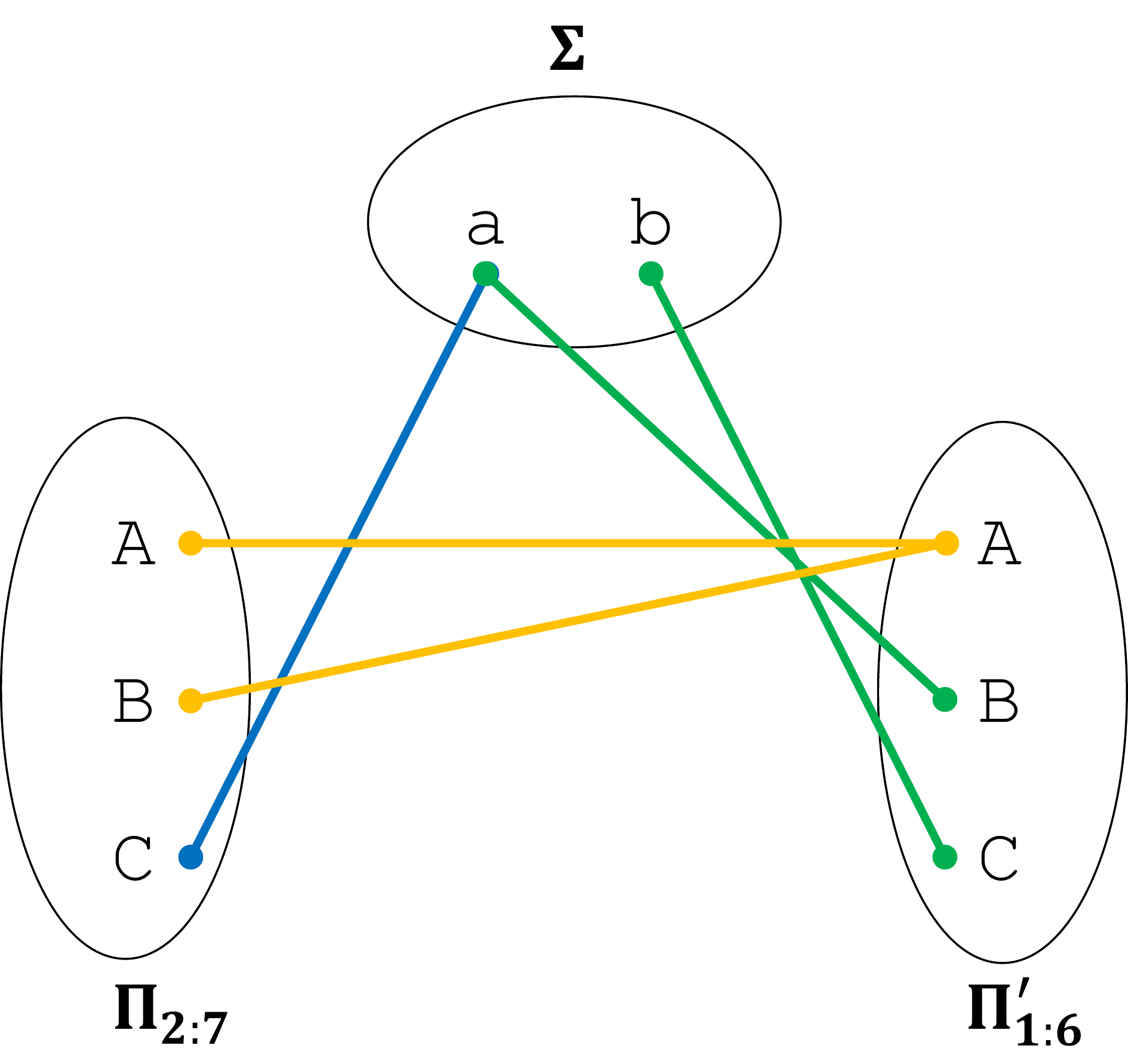}\\
	\ \ \ \small{(a) $(7,6)$-shifting graph}
	\end{minipage}\quad\quad
	\begin{minipage}[t]{0.36\hsize}
	\centering
	\includegraphics[width=\textwidth]{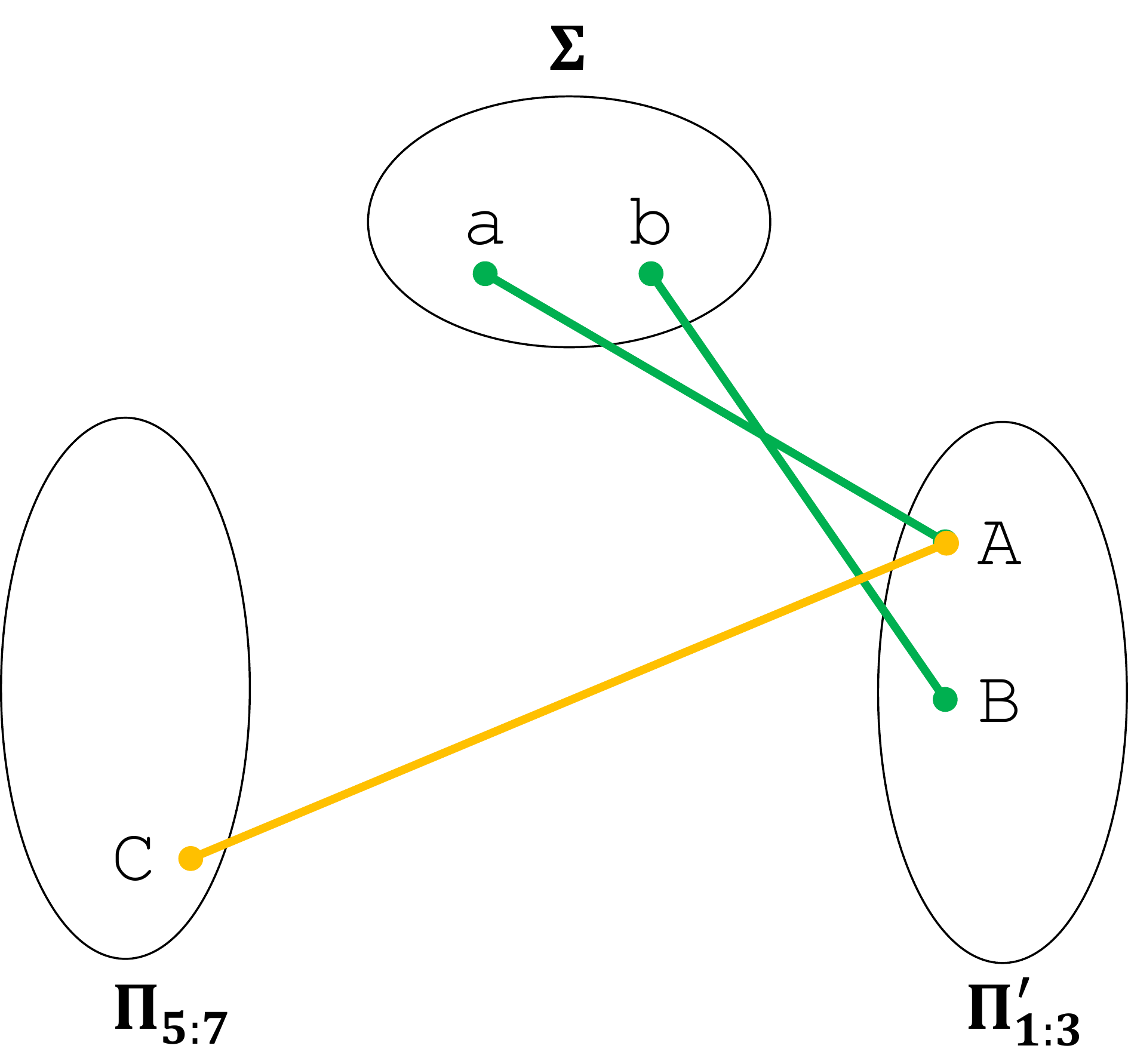}\\
	\ \ \  \small{(b) $(7,3)$-shifting graph}
	\end{minipage}
	\caption{The $(7,6)$-shifting graph (a) and $(7,3)$-shifting graph (b) on $P = \texttt{AABaaCbC}$, which corresponds to Fig.~\ref{fig:kmpexkmp}(i) and (ii).}
	\label{fig:shiftinggraph}
\end{figure}

\begin{definition}
	Let $\Pi'$ be a copy of $\Pi$ and $P'$ be obtained from $P$ by replacing every variable in $\Pi$ with its copy in $\Pi'$.
	For two numbers $k,j$ such that $0 \le j < k \le m$,
	the \emph{$(k,j)$-shifting graph} $G_{k,j} = (V_{k,j},E_{k,j})$ is defined by 
	\begin{align*}
	V_{k,j}&=\Sigma_P \cup \Pi_{P[k-j+1:k]} \cup \Pi'_{P'[1:j]}, \\
E_{k,j} &=\{\, (P[k-j+i], P'[i]) \mid 1 \le i \le j < k \text{ and } P[k-j+i] \neq P'[i] \,\} \,.
	\end{align*}

	We say that $G_{k,j}$ is \emph{invalid} if there are distinct $p,q \in \Sigma_P$ that belong to the same connected component.
	Otherwise, it is \emph{valid}.
\end{definition}
Note that $G_{k,0} = (\Sigma_{P},\emptyset)$ is valid for any $k$.
Figure~\ref{fig:shiftinggraph} shows the $(7,6)$-shifting and $(7,3)$-shifting graphs for $P=\mathtt{AABaaCbC}$ in Example~\ref{ex:exkmp}.
Using functions $\pi$ and $\pi'$ whose domains are $\textrm{dom}(\pi) = \Pi_{P[k-j+1:k]}$ and $\textrm{dom}(\pi') = \Pi_{P[1:j]}$, respectively, let us label each node $p \in \Sigma$, $x \in \Pi$, $x' \in \Pi'$ of $G_{k,j}$ with $p,\pi(x),\pi'(x)$, respectively.
Then $(\pi,\pi')$ is a $(k,j)$-shifting pair if and only if every node in each connected component has the same label.
Obviously $G_{k,j}$ is valid if and only if it admits a $(k,j)$-shifting function pair.

Thus, the resuming position should be $j+1$ for a mismatch at $k+1$ with a preceding function $\pi$ if and only if $j$ is the largest such that $G_{k,j}$ is valid and
\begin{itemize}
	\item[(a)] if $x \in \Pi$ and $p \in \Sigma$ are connected in $G_{k,j}$, then ${\pi}(x)=p$,
	\item[(b)] if $x \in \Pi$ and $y \in \Pi$ are connected in $G_{k,j}$, then ${\pi}(x)=\pi(y)$.
\end{itemize}
In that case, we have $\hat{\pi}'(P[1:j])=\hat{\pi}(P[k-j+1:k])$ for $\pi'$ determined by
\begin{itemize} 
	\item[(c)] $\pi'(x)=\hat{\pi}(y)$ if $x' \in \Pi'_{P[1:j]}$ and $y \in \Pi \cup \Sigma$ are connected.
\end{itemize}
We call the conditions (a) and (b) the \emph{$(k,j)$-preconditions} and (c) the \emph{$(k,j)$-postcondition}.
Note that every element in $\Pi'_{P'[1:j]}$ is connected to some element in $\Pi_{P[k-j+1:k]} \cup \Sigma_P$ in $G_{k,j}$ and thus $\pi'$ is well-defined.

\begin{remark}
The algorithm by du Mouza \etal{}~\cite{ref:du2007ppatternqueries} does not treat the condition induced by two nodes of distance more than 1 correctly.
For example, let us consider the pattern $P=\mathtt{AABaaCbC}$ in Example~\ref{ex:exkmp}.
For a text $T = \mtt{bbaaaabbb}$, the first mismatch occurs at $k=8$, where $\hat{\rho}(P[1:7]) = \mtt{bbaaaab}$ for $\rho(\mtt{A})=\mtt{b}$ and $\rho(\mtt{B})=\rho(\mtt{C})=\mtt{a}$.
To have $(\rho,\rho')$ a $(7,6)$-shifting pair for some $\rho'$, it must hold $\rho(\mathtt{A})=\rho(\mathtt{B})$.
That is, one can resume the comparison at position 6 only when the preceding function assigns the same constant to $\mathtt{A}$ and $\mathtt{B}$.
The preceding function $\rho$ in this case does not satisfy this constraint.
However, their algorithm performs this shift and reports that $P$ matches $T$ at position 2.
\end{remark}
To efficiently compute the failure function, our algorithm constructs another data structure instead of shifting graphs.
The \emph{shifting condition table} is a collection of functions $A_{k,j}: \Pi_{P[k-j+1:k]} \to \Pi_{P[k-j+1:k]} \cup \Sigma_{P}$ and $A'_{k,j}:\Pi'_{P'[1:j]} \to \Pi_{P[k-j+1:k]} \cup \Sigma_{P}$ for $1 \le j < k \le m$ such that $G_{k,j}$ is valid.
The functions $A_{k,j}$ can be used to quickly check the $(k,j)$-preconditions (a) and (b) and $A'_{k,j}$ is for the $(k,j)$-postcondition (c).
Those functions satisfy the following properties:
for each connected component $\alpha \subseteq V_{k,j}$, there is a representative $u_\alpha \in \alpha$ such that
\begin{itemize}
	\item if $\alpha \cap \Sigma \neq \emptyset$, then $u_\alpha \in \Sigma$,
	\item if $\alpha \cap \Sigma = \emptyset$, then $u_\alpha \in \Pi$,
	\item for all $x \in \alpha \cap \Pi$, then $A_{k,j}(x)=u_\alpha$,
	\item for all $x' \in \alpha \cap \Pi'$, then $A'_{k,j}(x') \in \alpha \cap (\Pi \cup \Sigma)$.
\end{itemize}
Recall that $G_{k-1,j-1}$ is a subgraph of $G_{k,j}$, where the difference is at most two nodes and one edge. 
Hence, we can compute $A_{k,j}$ and $A'_{k,j}$ in $O(|\Pi|)$ time from $A_{k-1,j-1}$ and $A'_{k-1,j-1}$ maintaining the inverse $U_{k,j}$ of $A_{k,j}$ whose domain is restricted to $\Pi$, i.e., $U_{k,j}(x) = \{\, y \in \Pi_{P[k-j+1:k]} \mid A_{k,j}(y) = x \,\}$ for $x \in \Pi_{P[k-j+1:k]}$.
Each set $U_{k,j}(x)$ can be implemented as a linked list.
The updating time $O(|\Pi|)$ is due to the size of $U_{k,j}$.
Moreover, when computing $A_{k,j}$ and $A'_{k,j}$, we can verify the validness of $G_{k,j}$.
A pseudo code for constructing the shifting condition table is shown as Algorithms~\ref{alg:fvctable} and \ref{alg:addconditionexf} in Appendix~\ref{sec:appendixalgorithms}.
\begin{lemma}
	The shifting condition table can be calculated in $O(|\Pi|m^2)$ time.
\end{lemma}
Suppose that we have a mismatch at position $k+1$ with a preceding function $\pi$.
By using the shifting condition table, a naive algorithm may compute the failure function in $O(k|\Pi|^2)$ time
by finding the largest $j$ such that $\pi$ satisfies the $(k,j)$-precondition and then compute a function $\pi'$ satisfying the $(k,j)$-postcondition with which we resume comparison at $j+1$.
The calculation of $\pi'$ can be done in $O(|\Pi|)$ time just by referring to the array $A'_{k,j}$.
We next discuss how to reduce the computational cost for finding $j$ by preparing an elaborated data structure in the preprocessing phase.

Du Mouza et al.~\cite{ref:du2007ppatternqueries} introduced a bitmap data structure concerning the precondition (a),
which can be constructed using $A_{k,j}$ in the shifting condition table as follows.
Here we extend the domain of $A_{k,j}$ to $\Pi$ by defining $A_{k,j}(x)=x$ for each $x \in \Pi \setminus \Pi_{P[k-j+1:k]}$.
\begin{definition}[\cite{ref:du2007ppatternqueries}]
	For every $0 \le j < k \le m$, $x \in \Pi$ and $p \in \Sigma_P$, we define
\begin{eqnarray*}
	r_{x, p}^{k}[j] &=& \begin{cases}
		0 & (\text{$G_{k,j}$ is invalid or $A_{k,j}(x) \in \Sigma \setminus \{p\}$})
\\		1 & (\text{otherwise})
	\end{cases}
\end{eqnarray*}
\end{definition}
\begin{lemma}[\cite{ref:du2007ppatternqueries}]\label{lem:precona}
	A preceding function $\pi$ satisfies the $(k,j)$-precondition (a)
	 if and only if
		$ \bigwedge_{x \in \Pi}{r_{x, \pi(x)}^{k}[j]}  = 1$.
\end{lemma}
We define a data structure corresponding to the $(k,j)$-precondition (b) as follows.
\begin{definition}
	For every $0 \le j < k \le m$ and $x,y \in \Pi$, define
	\begin{eqnarray*}
		s_{x, y}^{k}[j] &=& \begin{cases}
			0 & (\text{$G_{k,j}$ is invalid or $A_{k,j}(x) = y$}) \\
			1 & (\text{otherwise})
		\end{cases}
	\end{eqnarray*}
\end{definition}
\begin{lemma}\label{lem:preconb}
	A preceding function $\pi$ satisfies the $(k,j)$-precondition (b) if and only if 
		$ \bigwedge^{x,y \in \Pi}_{\pi(x) \neq \pi(y)} s_{x, y}^{k}[j]  = 1$\,.
\end{lemma}
Therefore, we should resume comparison at $j+1$ for the largest $j$ that satisfies the conditions of Lemmas~\ref{lem:precona} and~\ref{lem:preconb}.
To calculate such $j$ quickly, the preprocessing phase computes the following bit sequences.
For every $x \in \Pi$, $p \in \Sigma_P$ and $1 \le k \le m$, let $r_{x, p}^{k}$ be the concatenation of $r_{x, p}^{k}[j]$ in ascending order of $j$:
\begin{eqnarray*}
	r_{x, p}^{k} = r_{x, p}^{k}[0]\concat r_{x, p}^{k}[1]\concat \cdots \concat r_{x, p}^{k}[k-1]
\,,\end{eqnarray*}
and for every $x,y \in \Pi$ and $1 \le k \le m$, let
\begin{eqnarray*}
	s_{x, y}^{k} = s_{x, y}^{k}[0]\concat s_{x, y}^{k}[1]\concat \cdots \concat s_{x, y}^{k}[k-1]
\,.\end{eqnarray*}
Calculating $r_{x, p}^{k}$ and $s_{x, y}^{k}$ for all $x,y \in \Pi$, $p \in \Sigma_P$ and $1 \le k \le m$ in the preprocessing phase requires $O(|\Pi|(|\Sigma_{P}|+|\Pi|)m^2)$ time in total.
When a mismatch occurs at $k+1$ with a preceding function $\pi$, we compute
\[
	J = \bigwedge_{x \in \Pi} r_{x, \pi(x)}^{k}
	\wedge
	 	\bigwedge_{\substack{x,y \in \Pi \\ \pi(x) \neq \pi(y)}} s_{x, y}^{k} 
\,.\]
Then the desired $j$ is the right-most position of 1 in $J$.
This operation can be done in $O(\lceil \frac{m}{w} \rceil|\Pi|^2)$ time,
where $w$ denotes the word size of a machine.
That is, with $O(|\Pi|(|\Sigma_{P}|+|\Pi|)m^2)$ preprocessing time, the failure function can be computed in  $O(|\Pi|^2 \lceil \frac{m}{w} \rceil)$ time.
For most applications, we can assume that $m$ is smaller than the word size $w$, i.e. $\lceil \frac{m}{w} \rceil = 1$.

\begin{theorem}
	The \exfmatching{} problem can be solved in $O(|\Pi|^2 \lceil \frac{m}{w} \rceil n)$ time with $O(|\Pi|(|\Sigma_{P}|+|\Pi|)m^2)$ preprocessing time.
\end{theorem}

\subsection{Extended KMP Algorithm for \expmatch{}}

In this section, we consider the \expmatching{} problem.
We redefine the \emph{(mis)match} and \emph{failure function} in the same manner as described in the previous section except that all the functions are restricted to be injective.
We define $G_{k,j}$ exactly in the same manner as in the previous subsection.
However, the condition represented by that graph should be strengthened in accordance with the injection constraint on matching functions.
We say that $G_{k,j}$ is \emph{injectively valid} if for each $\Delta \in \{\Sigma , \Pi,\Pi'\}$, any distinct nodes from $\Delta$ are disconnected.
Otherwise, it is \emph{injectively invalid}. 
There is a $(k,j)$-shifting injection pair if and only if $G_{k,j}$ is injectively valid.

For $P=\mtt{AABaaCbC}$ in Example~\ref{ex:exkmp} (see Fig.~\ref{fig:shiftinggraph}), 
the $(7,6)$-shifting graph $G_{7,6}$ for $P=\mtt{AABaaCbC}$ is valid but injectively invalid, since $\mtt{A}$ and $\mtt{B}$ are connected.
On the other hand, $G_{7,3}$ is injectively valid.

In the PVC-matching, the condition that an injection pair $(\pi,\pi')$ to be $(k,j)$-shifting is described using the graph labeling by $(\pi,\pi')$ as follows:
\begin{itemize}
	\item two nodes are assigned the same label if and only if they are connected.
\end{itemize}
Under the assumption that $G_{k,j}$ is injectively valid, the $(k,j)$-precondition on a preceding function $\pi$ is given as
\begin{itemize}
	\item[(a)] if $x \in \Pi$ and $p \in \Sigma$ are connected, then ${\pi}(x)=p$,
	\item[(b')] if $x \in \Pi$ and $x' \in \Pi'$ are connected and $y' \in \Pi' \setminus \{x'\}$ and $p \in \Sigma$ are connected, then ${\pi}(x)\neq p$.
\end{itemize}
Since each connected component of an injectively valid shifting graph $G_{k,j}$ has at most 3 nodes, it is cheap to compute the function $F_{k,j}:V \to 2^{V_{k,j}}$ such that
$F_{k,j}(u) = \{\, v \in V_{k,j} \mid \text{$u$ and $v$ are connected in $G_{k,j}$}\,\}$.
Note that $F_{k,j}(u) = \emptyset$ if $u \notin \Pi_{P[k-j+1:k]}$.
Using $P[k],P[j]$, and $F_{k-1,j-1}$, one can decide whether $G_{k,j}$ is injectively valid and can compute $F_{k,j}$ (if $G_{k,j}$ is injectively valid) in constant time.

Suppose that we have a preceding function $\pi$ at position $k$.
By using the function $F_{k,j}$, a naive algorithm can compute the failure function in $O(k|\Pi|)$ time.
We define a bitmap $t_{x, p}^{k}[j]$ to check if $\pi$ satisfies preconditions (a) and (b').
\begin{definition}
	For every $0 \le j < k \le m$, $x \in \Pi$ and $p \in \Sigma_P$, we define
	\begin{eqnarray*}
		t_{x, p}^{k}[j] &=& \begin{cases}
			0 & (\text{$G_{k,j}$ is injectively invalid or }F_{k,j}(x) \cap \Sigma \nsubseteq \{p\} \\
			  &    \text{ or }  |F_{k,j}(x) \cap F_{k,j}(p) \cap \Pi'| = 2 )
\\			1 & (\text{otherwise})
		\end{cases}
	\end{eqnarray*}
\end{definition}
\begin{lemma}
The preceding function $\pi$ satisfies the $(k,j)$-preconditions (a) and (b')  if and only if
$ \bigwedge_{x \in \Pi}{t_{x, \pi(x)}^{k}[j]}  = 1$.
\end{lemma}
In the preprocessing phase, we calculate
\begin{eqnarray*}
	t_{x, p}^{k} = t_{x, p}^{k}[0]\concat t_{x, p}^{k}[1]\concat \cdots \concat t_{x, p}^{k}[k-1]
\end{eqnarray*}
for all $x \in \Pi$, $p \in \Sigma_P$ and $1 \le k \le m$, which requires $O(|\Pi||\Sigma_P|m^2)$ time.
When a mismatch occurs at $k+1$ with a function $\pi$, we compute
\[
	J = \bigwedge_{x \in \Pi} t_{x, \pi(x)}^{k}
\,\]
where the desired $j$ is the right-most position of 1 in $J$.
We resume comparison at $j+1$.
The calculation of the failure function can be done in $O(|\Pi|\lceil \frac{m}{w} \rceil)$ time, 
where $w$ denotes the word size of a machine.

\begin{theorem}
	The \expmatching{} problem can be solved in $O(|\Pi|\lceil \frac{m}{w} \rceil n)$ time with $O(|\Pi||\Sigma_{P}|m^2)$ preprocessing time.
\end{theorem}

\section{Concluding Remarks}

In this paper, we proposed efficient algorithms for the FVC-matching and PVC-matching problems.
The FVC-matching problem has been discussed by du Mouza \etal{}~\cite{ref:du2007ppatternqueries} as a generalization of the function matching problem,
while the PVC-matching problem is newly introduced in this paper, which can be seen as a generalization of the parameterized pattern matching problem.
We have fixed a flaw of the algorithm by du Mouza \etal{}\ for the FVC-matching problem.
There can be further variants of matching problems.
For example, one may think of a pattern with don't care symbols in addition to variables and constants.
This is not interesting when don't care symbols appear only in a pattern in function matching, since don't care symbols can be assumed to be distinct variables.
However, when imposing the injection condition on a matching function, don't care symbols play a different role from variables.
This generalization was tackled in~\cite{ref:yigarashi2017}.
We can consider an even more general problem by allowing texts to have variables, where two strings $P$ and $S$ are said to match if there is a function $\pi$ such that $\hat{\pi}(P)=\hat{\pi}(S)$.
This is a special case of the \emph{word equation problem}, where a string instead of a symbol can be substituted, and word equations are very difficult to solve in general.
Another interesting restriction of word equations may allow to use different substitutions on compared strings, i.e., $P$ and $S$ match if there are functions $\pi$ and $\rho$ such that $\hat{\pi}(P)=\hat{\rho}(S)$.
Those are interesting future work.

\bibliographystyle{splncs}
\bibliography{ref}

\newpage
\appendix
\section*{Appendix}
\section{Algorithms}
\label{sec:appendixalgorithms}

\begin{algorithm2e}[h]
	\DontPrintSemicolon
	\caption{The convolution-based algorithm for the \expmatching{} problem}
	\label{alg:fftbased}
	\KwIn{A string $P$ of length $m$, a string $T$ of length $n$}
	\KwResult{Every position $i$ such that $P$ \expmatches{} $T[i:i+m-1]$ }
	$\textit{result} \leftarrow \emptyset$ \\
	$G = \text{WildcardMatching}(T, P_{\dontcare})$ \tcc*[r]{Solve the wildcard matching problem}
	\ForEach{$b \in \Pi_{P}$}{
		$F_{b} \leftarrow T \conv \psi_{b}(P_{\Pi})$ \\
		$F'_{b} \leftarrow \TT \conv \psi_{b}(P_{\Pi})$ \\
		Let $c_{b}$ be the number of occurrences of $b$ in the pattern $P$
	}
	\For{$i \leftarrow 1$ \KwTo $n$}
	{
		$\textit{used} \leftarrow \emptyset$ \\
		$\textit{failed} \leftarrow false$ \\
		\ForEach{$b \in \Pi_{P}$} {
			$\textit{value} \leftarrow F_{b}[i]~/~c_{b}$ \\
			\If{$c_{b} \cdot F'_{b}[i] \ne (F_{b}[i])^2$ {\bf or}
			    $\textit{value} \in \textit{used}$}{
				$\textit{failed} \leftarrow true$ \\
				break
			}
			$\textit{used} \leftarrow \textit{used} \cup \{\textit{value}\}$
		}
		\If{$\textit{failed} = false$ {\bf and} $i \in G$}{
			$\textit{result} \leftarrow \textit{result} \cup \{i\}$			
		}
	}
	\KwRet{$\textit{result}$}
\end{algorithm2e}

\begin{algorithm2e}[h]
	\DontPrintSemicolon
	\caption{The shifting condition table construction algorithm for the \exfmatching{} problem}
	\label{alg:fvctable}
	\KwIn{A string $P$ of length $m$}
	\KwResult{The shifting condition table}
	Let \textit{table} be a 2d array of length $(m+1, m+1)$, where default value is \textit{NULL} \\
	Let $A_{0}: \Pi \to (\Sigma \cup \Pi)$ be the function s.t.\ $A_0[x]=x$ for all $x \in \Pi$ \\
	Let $A'_{0}: \Pi' \to (\Sigma \cup \Pi)$ be the function s.t.\ $A_0'[x']=x'$ for all $x' \in \Pi'$ \\
	Let $U_{0}:\Pi \to 2^\Pi$ be the function s.t.\ $U_0[x]=\{x\}$ for all $x \in \Pi$ \\
	$\textit{table}[1][0] \leftarrow (A_{0}, A'_{0}, U_{0})$ \tcc*[r]{Copy}
	\For{$k \leftarrow 2$ \KwTo $m+1$}
	{
		$\textit{table}[k][0] \leftarrow (A_{0}, A'_{0}, U_{0})$ \tcc*[r]{Copy}
		\For{$j \leftarrow 1$ \KwTo $k-1$} {
			\If{$\textit{table}[k-1][j-1] = \textit{NULL}$} {
				\Continue
			}
			$(A, A', U) \leftarrow \textit{table}[k-1][j-1]$ \tcc*[r]{Copy}
			$\alpha \leftarrow P[j]$ \\
			$\beta \leftarrow P[k]$ \\
			$\textit{valid} \leftarrow \textit{true}$ \\
			\If{$\alpha \in \Pi$} {
				\If{$A'[\alpha] \ne \textit{NULL}$ {\bf and} $A'[\alpha] \ne \beta$ } {
					\If{$\textit{AddCondition}(A, U, A'[\alpha], \beta) \ne \textit{VALID}$} {
						$\textit{valid} \leftarrow false$
					}
				}
				\Else {
					$A'[\alpha] \leftarrow \beta$
				}
			}
			\Else(\tcc*[f]{$\alpha \in \Sigma$}) {
				\If{$\textit{AddCondition}(A, U, \alpha, \beta) \ne \textit{VALID}$} {
					$\textit{valid} \leftarrow false$
				}
			}
			\If{$\textit{valid} = \textit{true}$} {
				$\textit{table}[k][j] \leftarrow (A, A', U)$
			}
		}
	}
	\KwRet{$\textit{table}$}
\end{algorithm2e}

\begin{algorithm2e}[h]
	\DontPrintSemicolon
	\caption{\textit{AddCondition} function for the \exfmatching{} problem}
	\label{alg:addconditionexf}
	\KwIn{Reference to function $A$ and array $U$ of $(k, j)$-shifting, symbols $a, b \in (\Pi \cup \Sigma)$}
	\KwResult{Whether $(k, j)$-shifting is valid or invalid after modifying $A$ and $U$}
	\If{$a \in \Pi$} {
		\If(\tcc*[f]{$a, b \in \Pi$}){$b \in \Pi$} { 		
			\If{$A[a] = A[b]$} {
				\KwRet{$\textit{VALID}$} \tcc*[r]{$A[a], A[b]$ are equal}
			}
		
			\If{$A[a] \in \Sigma$ {\bf and} $A[b] \in \Sigma$} {
				\tcc{$A[a], A[b]$ are connected to distinct symbols in $\Sigma$}
				\KwRet{$\textit{INVALID}$}
			}
			\ElseIf{$A[b] \in \Pi$} {
				\ForEach{$Z \in U[A[b]]$}{
					$A[Z] \leftarrow A[a]$
				}
				\tcc{Append linked list $U[A[a]]$ to the end of $U[A[b]]$}
				$U[A[a]] \leftarrow U[A[a]] \cup U[A[b]]$ \\
				$U[A[b]] \leftarrow \emptyset$ \tcc*[r]{Remove old linked list pointer}
			}
			\Else(\tcc*{$A[a] \in \Pi, A[b] \in \Sigma$}) {
				\ForEach{$Z \in U[A[a]]$}{
					$A[Z] \leftarrow A[b]$
				}
				$U[A[b]] \leftarrow U[A[b]] \cup U[A[a]]$ \\
				$U[A[a]] \leftarrow \emptyset$
			}
		}
		\Else(\tcc*[f]{$a \in \Pi, b \in \Sigma$}) {
			$\textit{root} \leftarrow A[a]$ \\
			\If{$root \in \Sigma \setminus \{b\}$} {
				\tcc{$a \in \Pi$ is already mapped to another symbol $\ne b$}
				\KwRet{\textit{INVALID}}
			}
			\ForEach{$Z \in U[root]$} {
				$A[Z] \leftarrow b$
			}
			$U[root] \leftarrow \emptyset$
		}
	}
	\Else {
		\If(\tcc*[f]{$a \in \Sigma, b \in \Pi$}){$b \in \Pi$} {
			\KwRet{\textit{AddCondition}(b, a)}
		}
		\Else(\tcc*[f]{$a, b \in \Sigma$}) {
			\If{$a \ne b$} {
				\KwRet{\textit{INVALID}}
			}
		}
	}
	\KwRet{\textit{VALID}}
\end{algorithm2e}

\end{document}